\documentclass[12pt]{article}

\title{\textbf{Idiosyncrasy as an explanation for power laws in nature\footnote{Paper accepted for publication in \textit{Trends in Mathematics}, Springer.}}}

\author{Salvador Pueyo\thanks{E-mail: spueyo@ic3.cat}\\\textit{\small{Institut Catal\`{a} de Ci\`{e}ncies del Clima (IC3), C/ Doctor Trueta 203, 08005 Barcelona,}}\\\textit{\small{Catalonia, Spain}}}

\date{}

\begin{document}

\maketitle

\begin{abstract}
\noindent Complex systems theory pays much attention to simple mechanisms producing nontrivial patterns, especially power laws. However, power laws with exponent $\tau \approx 1$ also result from complex mixtures of mechanisms that, in isolation, would not necessarily give this type of distribution. Probably, both paths to the power law are relevant in nature. The second gives a plausible explanation for some instances of power laws emerging in extremely complex systems, such as ecosystems.\\

\end{abstract}

\section{Idiosyncrasy in nature}
One central theme of complex systems theory is the identification of the mechanisms behind power laws in nature. Most attention has focused on simple mechanisms. Here I show that, in broad conditions, power laws with exponent close to $\tau = 1$ also arise from complex combinations of mechanisms that, in isolation, would not necessarily produce scale invariance. This contribution is a synthesis of papers by the author in journals of ecology and of climatology (the main ones being \cite{Pueyo2007b},\cite{Pueyo2012b}), which have, however, implications beyond these fields.

A fundamental difference between empirical research and simulations is the role of noise. Typically, empirical data result from the combination of the studied phenomenon with a myriad of other factors, whose joint effect is effectively random because of its high dimensionality. A major challenge of empirical research is separating these two components. In contrast, computers are unable to produce noise, with pseudorandom number generators being used to reproduce it imperfectly. In this aspect, simulation models give a biased view of nature.

In some empirical observations there might be no fundamental difference between \textit{signal} and \textit{noise}, if no single factor dominantes in the ocean of mechanisms contributing to each datum. This could well be the case e.g.\ for the \textit{species abundance distribution} (SAD), i.e.\ the frequency distribution $P(n)$ of the number $n$ of individuals belonging to each of the species in an ecological community or in a sample. Many simple mechanisms have been proposed to explain the shape of SADs \cite{McGill2007}, but naturalists are often skeptical about them. A huge number of elements affect the abundance of one given species (among them, its life cycle, its potential longevity and reproduction rate, the kinds of resources it consumes and the proportions in which they are needed, the types of places used e.g.\ for refuge or for laying eggs, its physiological response to a variety of environmental factors, the types of predators, parasites and infectious agents that can affect it and how they respond to its abundance, its capacity and patterns of movement, its propensity or not to form colonies or flocks or to more complex forms of cooperation, the peculiarities of its sexual reproduction if any, etc.). 

The variety of factors acting differently on different species suggests that each species could be ecologically \textit{idiosyncratic}, i.e.\ different from the rest in ways that cannot be described with a few words or a few parameter values. In fact, groups of related species will often share some features, but will differ in others. While biological species are not 100\% idiosyncratic, they appear to be so to a large extent. This is the basis of the \textit{idiosyncratic theory of biodiversity} by Pueyo et al.\ \cite{Pueyo2007b}. There could be many more instances of idiosyncrasy in other contexts.

This leads us to the need of identifying probability distributions that represent noise alone, with no signal. This could make us think of the normal or the lognormal, because of the central limit theorem. However, this theorem presumes a specific way of combining the causative factors (either additive or multiplicative) and these factors have some requirements in terms of mean and variance. We are interested in distributions that are even freer of information.

\section{Non-informative probability distributions}

\textit{Non-informative} distributions have been an object of controversy for more than one century, and caused statistics to split into several schools \cite{Kass1996}. While they interest us for other reasons, this controversy is motivated by their role in statistical inference. In Bayesian statistics, a variable or a parameter is assigned a probability distribution over a range of values even when only one can be correct (e.g.\ we can assign a distribution to a physical constant), expressing the plausibility of each possible value when the real one is unknown. For some data $D$, the PDF of a parameter $\theta$ is 
\begin{equation} 
f(\theta|D) \propto f(\theta)f(D|\theta),
\end{equation}
where $f(\theta)$ is the \textit{prior distribution}, meaning the distribution of probability (plausibility) that we would assign to $\theta$ if we did not know the data $D$. Equation 1 can be applied recursively, taking $f(\theta|D_1)$ as the prior probability density when introducing a second dataset $D_2$. Ideally, before introducing any data at all, the prior distribution should be \textit{non-informative}. In fact, in this (and in our) context, this adjective is attributed to distributions that may not be completely free from information, but give no information that is not contained in the enunciate of the problem. Many statisticians do not think that such a distribution exists, which is the reason why mainstream or \textit{frequentist} statistics avoids Eq.\ 1 by not assigning probabilities to parameters. Also among Bayesian statisticians there are a variety of positions about this issue \cite{Kass1996}.

Edwin T. Jaynes (best known for the maximum entropy principle \cite{Jaynes1957}) introduced a criterion known as \textit{group invariance} to determine non-informative distributions \cite{Jaynes1968}. It is especially appealing because it consists of a simple requirement of logical coherence. Its logic has similarities to that of relativity theory.

Consider e.g.\ the case in which we only know that the variable $x$ expresses a position. Let us introduce a new variable $y = x+ c$ (where $c$ is a constant), differing from $x$ only in a change of coordinates. Again, the only thing we know about $y$ is that it expresses a position, so we are forced to assign to it the same distribution as to $x$. The relation between the PDFs of reparameterized variables is well known: 
\begin{equation} 
f(y) = f(x)\left|\frac{dx}{dy}\right|.
\end{equation}
The only distribution that leaves the PDF unchanged when introducing $y = x+ c$ into Eq.\ 2 is the uniform, in agreement with intuition. 

In contrast to the case of positions, for other variables and parameters there is no ambiguity in the origin of coordinates, because \textit{zero} has a well-defined meaning. However, in some of these cases a change of scale $y = cx$ (e.g.\ a change of units \cite{Baker2007}) leaves the enunciate of the problem unchanged. Then, the distribution that remains invariant under Eq.\ 2 is a power law with exponent $\tau = 1$:
\begin{equation} 
f(x) \propto x^{-1}.
\end{equation} 

For example, the author proposed to apply this criterion in the inference of climate sensitivity $S$ \cite{Pueyo2012b}, which quantifies the roughly linear relation between climate forcing $\Delta F$ (e.g.\ due to emitted CO$_{2}$) and global warming $\Delta T$ after reaching a steady state: 
\begin{equation} 
\Delta T = S \Delta F.
\end{equation} 
Note that, if our distribution is to encode only the form of Eq.\ 4, it should not change when inverting it, because it follows from Eq.\ 4 that 
\[
\Delta F = \lambda \Delta T, 
\]
where $\lambda$ is the climate feedback parameter $\lambda = S^{-1}$. This is true whenever the PDF of $z = \ln(S)$ (i.e.\ $z = -\ln(\lambda)$) is symmetric about $z = 0$, but $z = 0$ corresponds to $S = 1$, whose physical meaning depends on the units. As units are arbitrary (and do not alter the form of Eq.\ 4), the PDF of $z$ should be symmetric not only about $z = 0$ but also about any other value of $z$, i.e.\ the distribution of $z$ should be uniform, which translates to the power law in Eq.\ 3 when expressed in terms of $S$ (or of $\lambda$) \cite{Pueyo2012b}: 
\begin{equation} 
f(S) \propto S^{-1}.
\end{equation}

\section{Frequency distributions with little information}

The two previous sections give us the two key ingredients for the theoretical framework presented in this paper: idiosyncrasy (Sect. 1) and Jaynes' group invariance criterion (Sect. 2). Besides using the distributions obtained with this criterion as a tool for inference, Jaynes \cite{Jaynes1973} noted that they could have a translation to frequency distributions in some cases, but he does not seem to have developed the notion of idiosyncrasy as a condition for this correspondence.

We obtain the prior in Eq.\ 3 as an intermediate step to estimate a parameter with a single value, $S$. However, the parameters of other equations with the same shape as Eq.\ 4 can take more than one value. Consider e.g.\ Fourier's equation of thermal conductivity 
\[
\mathbf{q} = -k\nabla T.
\]
Different materials have different $k$, so, unlike $S$, we can obtain an empirical frequency distribution for $k$. While it is highly plausible that the abundances of different biological species can be considered idiosyncratic because of the variety and complexity of factors determining them, this is not so clear for the case of the thermal conductivities of different materials. However, starting from this hypothesis, the empirical PDF of $k$ in 189 solid materials was analyzed by the author (\cite{Pueyo2012b}, sect. 4.3) and did result into a distribution very close to Eq.\ 3 in a broad range. Note that this is an instance of empirical power law that can be hardly interpreted as the result of any process of self-organization. Grandison and Morris \cite{Grandison2008} also found a figure close to Eq.\ 3 for the frequency distribution of biochemical pathway kinetic rate parameters. 

In the case of SADs, if species are fully idiosyncratic, knowing $n$ for an unidentified species should give us no information about anything else. Since an SAD contains no reference to e.g.\ location, whichever distribution we choose will remain invariant when changing location. However, for most  types of distribution, the value of $n$ would give us some clue whether it refers to a whole continent or to the researcher's backyard. There is only one distribution for which the SAD does not change when changing spatial scale. Even though $n$ is a discrete variable, Pueyo et al.\ \cite{Pueyo2007b} proved that this distribution is the direct analogue of Eq.\ 3: 
\begin{equation} 
P(n) \propto n^{-1}.
\end{equation} 
However, if, as usual, we are dealing with sets of species with a characteristic size (e.g.\ insects), Eq.\ 6 cannot be our final answer, because the total number of individuals of all the species in the community is limited for physical reasons, while Eq.\ 6 would imply that some species would exceed this number of individuals, whichever it is. 

There is a well-known method to obtain a distribution with the least possible information subject to some specified constraints: the maximum entropy formalism (MaxEnt), also due to Jaynes \cite{Jaynes1957}. Even though Jaynes \cite{Jaynes2003} was clear about this point, there is little awareness that the non-informative distribution is a needed input for MaxEnt.

In the case of SADs, if we start from Eq.\ 6 and we apply MaxEnt to constrain the mean $\overline{n}$, we obtain a power law with $\tau = 1$ and an exponential bending function \cite{Pueyo2007b}: 
\begin{equation} 
P(n) \propto n^{-1} \exp ^{-\phi n}.
\end{equation}
Equation 7 corresponds to the logseries distribution, which was one of the first ever used to fit an empirical SAD \cite{Fisher1943} and is still widely applied. While some empirical SADs are well fitted by Eq.\ 7, others display small departures, which is not surprising because species are not fully idiosyncratic. The author showed that sharing resources or predators among species does not, in principle, alter Eq.\ 7, but similarities in the shape of the equations describing their population dynamics can have an effect \cite{Pueyo2007b}. However, using Taylor series, Pueyo \cite{Pueyo2006b} also found that most empirical SADs can be explained from small perturbations around Eq.\ 7. The simplest deviation consists of a truncated power law with an exponent slightly different from $\tau=1$.

\section{Conclusions}

Probably, there are many instances in which an empirical power law results from some specific mechanism that can be reproduced by simple models, e.g.\ self-organized criticality \cite{Pruessner2012}. However, some hypercomplex systems can produce observables that are effectively \textit{noise}, and, in many cases, noise implies a distribution close to a power law with exponent $\tau = 1$. It is essential to take this option into account when studying complex systems in the real world. 

\section*{Acknowledgements}

I am grateful to \'{A}lvaro Corral for his suggestion to present this paper at the CRM \& Imperial College Workshop on Complex Systems, 12-13 April 2013. It was presented in that event and is accepted for publication in \textit{Trends in Mathematics}, Springer.

\bibliographystyle{unsrt}

\end{document}